\documentclass{article}
\usepackage{amsmath, amssymb, verbatim}
\usepackage{subfig}
\usepackage[dvips]{graphicx}
\usepackage[dvips]{color}

\title{Phase Transition and Anisotropic Deformations of Neutron Star Matter}
\author{Susan Nelmes\footnote{email:s.g.nelmes@durham.ac.uk}
\quad and \quad 
Bernard M. A. G. Piette\footnote{email:b.m.a.g.piette@durham.ac.uk}}

\begin{document}

\maketitle

\begin{center}
                                                                             
{\em Department of Mathematical Sciences,
University of Durham,
Science Laboratories,
South Road,
Durham. DH1 3LE.
UK}
  
\end{center}

\abstract 
The Skyrme model is a low energy, effective
field theory for QCD which when coupled to a gravitational field
provides an ideal semi-classical model to describe neutron stars. We use the
Skyrme crystal solution composed of a lattice of $\alpha$-like particles as a
building block to construct minimum energy neutron star
configurations, allowing the crystal to be strained
anisotropically. We find that below 1.49 solar masses the stars' crystal 
deforms isotropically and that above this critical mass, it undergoes
anisotropic strain. We then find that the maximum mass allowed for 
a neutron star is 1.90 solar masses, in close agreement with a
recent observation of the most massive neutron star yet found.
The radii of the computed solutions also match the experimentally estimated 
values of approximately 10km.
\newpage

\section{Introduction}
Neutron stars are stars that have collapsed under intense self gravitational 
pressure to the point where all electrons are squeezed into nuclei, 
hence forming a large cluster of neutrons with a typical radius of about 10km. 
A neutron star can thus be seen as 
a gigantic nuclei that is electrically neutral but is strongly affected by 
the gravitational field that it generates. For lack of a unified theory of
strong interactions and gravity one has to resort to finding an approximate 
theory that allows us to describe such a system.

One such theory is the Skyrme model. Originally proposed by
Skyrme in 1961 \cite{Skyrme:1961vq}, \cite{Skyrme:1962vh}
as a nonlinear theory of pions to describe strong interactions, it was later
shown by Witten \cite{Witten:1983tw} to be
an approximate, low energy, effective field theory for QCD which
becomes more exact as the number of quark colours becomes large.

Each solution of the Skyrme model is characterised by an integer valued
topological charge which can be identified with the baryon number $B$. 
The simplest solution, $B=1$, is made out of a so-called Skyrmion and 
corresponds to a proton or neutron. 
At the semi-classical level, 
the Skyrme model does not distinguish between a neutron and a proton. 
Moreover, as the model does not include the electroweak interaction, all
Skyrmions are electrically neutral. 

The $B=1$ solution of the Skyrme model is the only exact stable 
solution that can be computed easily \cite{Skyrme:1961vq}.
Solutions with larger values of $B$ can only be computed numerically 
\cite{Braaten:1990},\cite{Battye:1997wf} and these solutions have been shown
to successfully describe various nuclei and their 
properties \cite{Battye:2009ad}.

Moreover, one can also compute crystal-like solutions made out of an 
infinite number of Skyrmions. In particular, it has been shown that the 
Skyrme solution with the lowest energy per Skyrmion corresponds to a cubic 
lattice where each lattice unit has a topological charge $B=4$
\cite{Castillejo1989801}.
These solutions can thus be seen as a crystal of $\alpha$ particles.

This lattice of Skyrmions looks thus as the best building block to describe
a neutron star as its has the lowest possible energy per baryon. 
Yet, one must first
estimate if a star could instead be made out of a liquid or gas of Skyrmions.
The temperature of a neutron star, a few years after its creation,
cools down to an approximate temperature of about 
$100\mbox{eV}\approx 10^6K$ \cite{baldo1999nuclear}. While this looks 
like a very high temperature compared to the binding energy of an electron 
around a nucleus, this energy is quite small from a nuclear point of view. 
Indeed, the lowest excited state of an $\alpha$ particle, for example, 
is $23.3$MeV \cite{Tilley19921} and the lowest vibration mode of a $B=4$
Skyrmion is of the order of $100\mbox{MeV}$ \cite{Barnes:1997},\cite{Lin:2008}.
Even under intense gravitational energy, Walhout \cite{Walhout:1987ce}
showed that the excitation energy of a lattice of $B=1$ Skyrmions is also of
the order of $100\mbox{MeV}$. This points out that the neutron 
star will be in a solid phase rather than a liquid or a gas and that the 
thermal energy will only excite acoustic phonon modes. 
It is thus natural to model a neutron star as a lattice 
of $B=4$ Skyrmions. 

Before we proceed we must also question the possibility of having  
an atmosphere around the star, and to estimate its height if it turns out not 
to be small. At the surface of a neutron star twice the mass of the sun, 
the gravitational acceleration is $g\approx2.6\times 10^{12}\mbox{ms}^{-2}$.
It is then easy to compute that the average height that an $\alpha$ particle
with a thermal energy of $100\mbox{eV}$ will be able to jump is of the order 
of $1mm$, {\it i.e.} much smaller than the radius of the star. We can thus 
consider that such an atmosphere is extremely thin and assume in our model 
that the neutron star is fully made out of a solid.

Having established that we can use a Skyrme crystal as the building block to 
describe a neutron star we will proceed as follows. First of all, we will use 
the equations of state computed by Castillejo {\it et al.} 
\cite{Castillejo1989801} 
for the $B=4$ crystal when the lattice is deformed asymmetrically. 
Following Walhout \cite{Walhout:1987ce} we will then use
a Tolman-Oppenheimer-Volkoff (TOV) equation \cite{Tolman:1939jz},
\cite{Oppenheimer:1939ne}, generalising it to allow for matter to be 
anisotropic \cite{1974ApJ...188..657B}. The TOV equation describes the static 
equilibrium between matter forces within a solid or fluid and the 
gravitational forces self-generated by the matter for a spherically 
symmetric body. 

Combining the TOV equation with the equations of state of
the Skyrme crystal, we will be able to find 
configurations that are spherically symmetric distributions of 
anisotropically deformed matter in static
equilibrium and so are suitable to model neutron stars.
Solving these equations numerically for large 
stars we will show that below a critical mass of $1.49$ solar masses 
($M_\odot=1.98892\times10^{30}$kg) all
neutron/Skyrmion stars are made out of an isotropically strained crystal. 
We will then 
show that at this critical mass, there is a phase transition and that heavier 
stars are made out of an anisotropically deformed crystal that is 
less strained radially than tangentially. We will also show that these 
stars can have a mass of up to $1.90M_\odot$.   
Finally, we will investigate the impact of adding a mass term to the Skyrme 
model and describe what happens to a star when its mass is increased above its 
maximum value. 

Using Skyrmions to model neutron stars is not new and has been performed
previously in several ways. First of all, Walhout used a lattice of $B=1$ 
Skyrmions \cite{Walhout:1987ce} to describe a neutron star. He then improved
his results by considering a lattice of $B=4$ Skyrmions \cite{Walhout:1990va}. 
In both cases he assumed an 
isotropic compression of the lattice, assuming a gas-like phase, and he
used numerical solutions of the model to estimate the stress tensor. 
The maximum mass he obtained for the neutron star was $2.57M_\odot$.
Later, Jaikumar and Ouyed \cite{Jaikumar:2005gq} considered the equation of 
state for a neutron star based on a Skyrme fluid and obtained a maximum 
mass of $3.6M_\odot$. The main difference between these 2 approaches and ours 
is that they assumed an isotropic fluid of Skyrmions whereas we
consider a solid crystal allowed to deform anisotropically, {\it i.e.} 
be compressed differently in the radial and tangential directions of the star. 
In our previous papers \cite{Piette:2007wd},
\cite{Nelmes:2011zz}, we computed minimal energy Skyrmion stars
made out of layers of 2 dimensional Skyrme lattices. This allowed us to use 
the rational map ansatz \cite{Houghton:1997kg} to minimise the energy directly
but resulted in relatively small stars with a maximum mass of 
$0.574M_\odot$. This was 
mainly due to the fact that the field transition between the different layers
in our ansatz over estimated the energy of the configuration and that the 
energy per baryon in each layer of the ansatz was also larger than that of 
the crystal of $B=4$ Skyrmions.

\section{Skyrme Crystals}

The Skyrme model \cite{Skyrme:1961vq}, \cite{Skyrme:1962vh} is described by 
the Lagrangian
\begin{equation}
\label{Lagrangian}
\mathcal{L}_{Sk} = \frac{F_{\pi} ^2}{16} 
  {\rm{Tr}} (\nabla_{\mu} U \nabla^{\mu} U^{-1})
 + \frac{1}{32 e^2} {\rm{Tr}} [(\nabla_{\mu} U) U^{-1}, 
 (\nabla_{\nu} U) U^{-1}]^2 + \frac{m_{\pi}^2F_{\pi}^2}{8}
 {\rm{Tr}}(U-1),
\end{equation}
where here $U$, the Skyrme field, is an $SU(2)$ matrix and $F_{\pi}$, $e$ 
and $m_{\pi}$ are the pion decay constant, the Skyrme coupling and the pion 
mass term respectively. In the Lagrangian (\ref{Lagrangian}) the $\nabla$
are ordinary partial derivatives in the absence of a gravitational field and 
become covarient derivatives when the Skyrme field is coupled to  
gravity. The results summarised in this section all relate to the pure 
Skyrme model without gravity.

The Skyrme field is a map from $\mathbb{R}^{3}$ to $S^3$, the group
manifold of $SU(2)$, but finite energy considerations imply that the
field at spatial infinity should map to the same point, meaning the
Skyrme field is a map between two three-spheres. Such maps fall into
homotopy classes indexed by an integer, known as the topological
charge, which is interpreted as the baryon number, $B$. The
topological soliton solutions, known as Skyrmions, are identified as
baryons with an $\alpha$ particle described by a $B=4$ Skyrmion
solution.\par

Here we will be considering the zero pion mass case 
where $m_{\pi}=0$ with section $4.3$ describing the effects of its 
inclusion. The two other Skyrme parameters, $F_\pi$ and $e$ can be obtained 
in different ways. Skyrme first evaluated them by taking the experimental
value of the pion decay constant $F_\pi=186\mbox{MeV}$ and then 
fitting the mass of a Skyrmion to that of a proton and obtained $e=4.84$. 
Later Adkins, Nappi and Witten \cite{Adkins:1983ya} 
quantised the $B=1$ Skyrmion to fit 
the parameter values to the mass of the nucleon and the delta excitation and 
obtained $F_\pi=129\mbox{MeV}$ and $e=5.45$. These later values were 
the ones used by Castillejo {\it et al.} \cite{Castillejo1989801} to compute 
the energy of the deformed $B=4$ crystal and we will thus use them too.

The solution of the Skyrme model with the lowest energy per baryon 
has been shown to be a face-centred cubic (fcc) lattice of
Skyrmions \cite{Kugler1988491}, \cite{Castillejo1989801}.
Each unit cell is a cube of side length $a$ with a baryon
number of $B=4$ and can therefore be considered as an
$\alpha$ particle. 
In the context of a neutron star, we will be able to interpret each $B=4$ 
crystal component as being 4 neutrons as the Skyrme model does not
distinguish between neutrons and protons.
\par


Castillejo \emph{et al.} \cite{Castillejo1989801} also investigated
the energy of dense Skyrmion crystals where the configuration
was not a face-centred cubic lattice but rather a lattice where the
aspect ratio of the unit cell, $B=4$ $\alpha$ particle, of side $a$ was
altered so that it becomes rectangular with aspect ratio $r^3$. This
means that in the $x$ and $y$ directions the lattice size becomes
$ra$ and in the $z$ direction, $a/r^2$. As in
Castillejo \emph{et al.} we use the measure $p=r-1/r$ to describe the
deviation away from the face-centred cubic lattice symmetries which
have $p=0$.


The numerical solutions found in \cite{Castillejo1989801} provide an
equation for the dependence of the energy of a single Skyrmion, $E(L,p)$, 
on its size, $L=n ^{-1/3}$, where $n$ is the Skyrmion number density, and its
aspect ratio measure, $p$. 

\begin{equation}
\label{Energy}
 E(L,p)=E_{p=0}(L)+E_0[\alpha(L)p^2+\beta(L)p^3+\gamma(L)p^4+\delta(L)p^5
  +...],
\end{equation}
 where the coefficients are given by
\begin{eqnarray}
 E_{p=0}(L)&=&E_0\left[0.474\left(\frac{L}{L_0}+\frac{L_0}{L}\right)
 +0.0515\right],\\
\alpha(L)&=&0.649-0.487\frac{L}{L_0}+0.089\frac{L_0}{L},\\
\beta(L)&=&0.300+0.006\frac{L}{L_0}-0.119\frac{L_0}{L},\\
\gamma(L)&=&-1.64+0.78\frac{L}{L_0}+0.71\frac{L_0}{L},\\
\delta(L)&=&0.53-0.55\frac{L}{L_0}.
\end{eqnarray}
Here $E_0=727.4$MeV and $L_0=1.666\times10^{-15}$m. The
equation can be extended to include lower densities
\cite{Castillejo1989801} but they are not of interest here where we
are only considering densities higher than the minimal energy
crystal. Notice that for any value of $L$ the minimum energy
occurs at the face-centred cubic lattice configuration, $p=0$,
and the global minimum is reached for $L=L_0$.\par 

\section{TOV Equation for Skyrmion Stars} 
Using equation \eqref{Energy} relating the energy of a Skyrmion to its
size and aspect ratio we will now investigate how one can describe a
neutron star using a Skyrme crystal and how this crystal is
deformed under the high gravitational field it experiences.\par

In our numerical work we denote $\lambda_r$ as the Skyrmion length in the 
radial direction of the star and  $\lambda_t$ as the Skyrmion length in the
tangential direction. These parameters and the parameters $L$ and $p$ 
used in \eqref{Energy} are related as follows
\begin{equation}
 L=(\lambda_r\lambda_t\lambda_t)^{\frac{1}{3}},\ {\rm{and}}\ 
 p=\left(\frac{\lambda_t}{\lambda_r}\right)^\frac{1}{3}
  -\left(\frac{\lambda_r}{\lambda_t}\right)^\frac{1}{3}.
\label{eq_L}
\end{equation}
\par

To construct a neutron star we consider a
spherically symmetric distribution of matter in static equilibrium
with a stress tensor that is in general locally anisotropic. Spherical
symmetry demands that the stress tensor, $T_\nu^\mu$, is diagonal and
that all the components are a function of the radial coordinate
only. We denote this stress tensor as 
\begin{equation}
 T_\nu^\mu={\rm diag}(\rho(r),p_r(r),p_\theta(r),p_\phi(r)),
\label{eq_Tmunu}
\end{equation}
and consider that, again due to spherical symmetry, $p_\theta(r) =
p_\phi(r)$ which we will denote by $p_t(r)=p_\theta(r) =
p_\phi(r)$. The quantities $p_r(r)$ and $p_t(r)$ describe the stresses
in the radial and tangential directions respectively while the
quantity $\rho(r)$ is the mass density.\par 

A generalised TOV equation
\cite{Tolman:1939jz}, \cite{Oppenheimer:1939ne} to describe a
spherically symmetric star composed of anisotropically deformed matter 
in static equilibrium has been studied previously
\cite{1974ApJ...188..657B} and we summarise it now. 

The metric for the static spherically symmetric distribution of matter
can be written in Schwarzschild coordinates as 
\begin{equation}
 ds^2=e^{\nu(r)} dt^2-e^{\lambda(r)} dr^2-r^2d\theta^2
      -r^2\sin^2\theta d\phi^2,
\label{metric}
\end{equation}
where $e^{\nu(r)}$ and $e^{\lambda(r)}$ are functions of the radial
coordinate that need to be determined. 
The combination of this metric and the matter
distribution, described by the stress tensor (\ref{eq_Tmunu}), 
must be a solution of Einstein's equations
\begin{equation}
 G_{ab}=R_{ab}-\frac{1}{2}Rg_{ab}=8\pi T_{ab},
\end{equation}
where we have set $G=c=1$. After calculating the Ricci tensor and Ricci 
scalar from the metric we find
\begin{eqnarray}
e^{\lambda} \left( \frac{{\lambda}^{\prime}}{r} - \frac{1}{r^2} \right) 
  + \frac{1}{r^2} &=& 8{\pi}{\rho}\label{EE1}\\
 e^{-\lambda} \left( \frac{{\nu}^{\prime}}{r}+ \frac{1}{r^2} \right) 
  - \frac{1}{r^2} &=& 8{\pi}p_{r}\label{EE2}\\
e^{-\lambda} \left( \frac{1}{2} {\nu}^{\prime \prime} 
  - \frac{1}{4}{\lambda}^{\prime}{\nu}^{\prime} 
  + \frac{1}{4} \left({\nu}^{\prime} \right)^{2} 
  +  \frac{\left({\nu}^{\prime} -{\lambda}^{\prime} \right)}{2r} \right) 
&=& 8{\pi}p_{t}~.\label{EE3}
\end{eqnarray}
Equation \eqref{EE1} can be rewritten as
\begin{equation}
 (re^{-\lambda})'=1-8\pi \rho r^2
\end{equation}
and integrated to give 
\begin{equation}
 e^{-\lambda}=1-\frac{2m}{r}
\label{EE4}
\end{equation}
where $m = m(r)$ is defined as the gravitational mass contained within the 
radius $r$ and can be calculated by
\begin{equation}
 m=\int^r_04\pi r^2\rho dr.
\label{eq_m}
\end{equation}
We can now substitute equation \eqref{EE4} for $e^{-\lambda}$ into 
equation \eqref{EE2} to find
\begin{equation}
 \frac{1}{2}\nu'=\frac{m+4\pi r^3 p_r}{r(r-2m)}.
\label{eq_nuprimeTOV}
\end{equation}\par
The generalised TOV equation that we will use to find 
suitable neutron star configurations can now be obtained by differentiating 
equation \eqref{EE2} with respect to $r$ and adding it to equation 
\eqref{EE3} to find
\begin{equation}
 \frac{dp_{r}}{dr}  = -(\rho + p_{r})\frac{{\nu}^{\prime}}{2} 
 + \frac{2}{r}(p_{t} - p_{r})~.
\label{TOV}
\end{equation}
Now, substituting (\ref{eq_nuprimeTOV}) into (\ref{TOV}), we get
\begin{equation}
 \frac{dp_{r}}{dr}  = -(\rho + p_{r})\frac{m+4\pi r^3 p_r}{r(r-2m)}
 + \frac{2}{r}(p_{t} - p_{r})~.
\label{eq_mainTOV}
\end{equation}\par

For this generalised TOV equation to be
solvable two equations of state need to be specified, ${p_{r} =
p_{r}(\rho)}$ and ${p_{t} = p_{t}(\rho)}$, where, as argued above, 
we are able to use a zero temperature assumption.

We also need to specify appropriate boundary conditions. 
First, we must require
that the solution is regular at the origin and impose that $m(r)
\rightarrow 0$ as $r \rightarrow 0$. Then $p_{r}$ must be finite at the
centre of the star implying that ${\nu}^{\prime} \rightarrow 0$ as $
r \rightarrow 0$. Moreover, the gradient $dp_{r}/dr$ must be finite at 
the origin too and so $(p_{t} - p_{r})$ must vanish at least as rapidly 
as $r$ when $r\rightarrow 0$. This implies that we need to impose the boundary
condition $p_{t}=p_{r}$ at the centre of the star.\par 

The radius of the star, $R$, is determined by the condition $p_{r}(R)
= 0$ as the radial stress for the Skyrmions on the surface of the star
will be negligibly small. The equations, however, do not impose that
$p_{t}(R)$ vanishes at the surface. One should also point out that
physically relevant solutions will all have $p_{r} ,p_{t} \geq 0 $ for
$r \leq R$. We note that an exterior vacuum Schwarzschild metric can
always be matched to our metric for the interior of the star across
the boundary $r=R$ as long as $p_{r}(R) = 0$, even though $p_{t}(R)$
and $\rho_{r}(R)$ may be discontinuous, implying that the star can
have a sharp edge, as expected from a solid rather than gaseous
star.\par  

As we are considering Skyrmion matter at zero temperature the
equations of state that will be used in finding suitable neutron star
configurations can be calculated from equation \eqref{Energy} which
depends on the lattice scale $L$, and aspect ratio, $p$, which are both
functions of the radial distance form the centre of the star, $r$.
From the theory of elasticity we then find that the radial and the tangential 
stresses are related to the energy per Skryrmion, Eq (\ref{Energy}), as follows
\begin{equation}
 p_r=-\frac{1}{\lambda_t^2}\frac{\partial E}{\partial \lambda_r},\ 
{\rm{and}}\ p_t=-\frac{1}{\lambda_r}\frac{\partial E}{\partial \lambda_t^2}.
\label{EOS}
\end{equation}

Using the generalised TOV equation \eqref{eq_mainTOV} and the 
two equations of state \eqref{EOS}, a minimum energy configuration for 
various values of the total baryon number can be calculated numerically. 
The minimum energy configuration is defined as the minimum value of the 
gravitational mass, $M_G$,
\begin{equation}
 M_G=m(R)=m(\infty)=\int_0^R4\pi r^2\rho dr,
\label{eq_MG}
\end{equation}
where $R$ is the total radius of the star and
\begin{equation}
 \rho = \frac{E}{\lambda_r \lambda_t^2 c^2}.
\label{eq_rho}
\end{equation}

We now need to minimise $M_G$ as a function of $\lambda_r$ and $\lambda_t$ 
which both depend on $r$. To achieve this, we will first assume a profile
for $\lambda_t(r)$ and compute $M_G$ for this profile as described below.
We will then determine the configuration of the neutron star, with a specific 
baryon charge, by minimising $M_G$ over the field $\lambda_t$.
This can be easily done using the simulated annealing algorithm. 

To compute $M_G$ we notice that 
at the origin, one can use (\ref{EOS}) to determine  $p_r(0)$ and $p_t(0)$
from the initial values of $\lambda_r(0)$ and $\lambda_t(0)$.
Then the integration steps can be performed as follows.
Knowing $\lambda_r(r)$ and $\lambda_t(r)$ one computes $\rho(r)$ 
using (\ref{eq_rho}) and $m(r)$ using (\ref{eq_m}). 
Then, knowing $p_r(r)$, $p_t(r)$, $\rho(r)$ and $m(r)$  
one can integrate (\ref{eq_mainTOV}) by one step to determine $p_r(r+dr)$. 
One can then use (\ref{EOS}) to determine $\lambda_r(r+dr)$ and as 
the profile for $\lambda_t(r)$ is fixed, one can proceed
with the next integration step.

One then integrates (\ref{eq_mainTOV}) up to the radius $R$ for which 
$p_r(R)=0$; this sets the radius of the star. 
In our integration, we used a radial step of $50$m.

One must then evaluate the total baryon charge of the star using
\begin{equation}
 B = \int_0^R \frac{4\pi r^2 n(r)}{(1-\frac{2 G m}{c^2 r})^{1/2}}dr
\label{eq_B}
\end{equation}
where 
\begin{equation}
 n(r) = \frac{1}{\lambda_r(r)\lambda_t(r)^2}
\end{equation}
and rescale $\lambda_t$ to restore the baryon number to the desired value.
One then repeats the integration procedures until the baryons charge reaches 
the correct value without needing any rescaling.

\section{Results}

\subsection{Stars Made of Isotropically Deformed Skyrme Crystal} 
We found that up to
a baryon number of $2.61\times10^{57}$, equivalent to $1.49M_\odot$, 
the minimum energy configurations are all
composed of Skyrme crystals that are isotropically deformed, with
$\lambda_t(r)=\lambda_r(r)$ across the whole radius of the star. It
can be shown that this indeed has to be the case as we can prove that
if it is possible to find an isotropic Skyrme crystal solution then that
solution will be the minimum energy configuration. Such isotropic
solutions can only be found up to a baryon number of
$2.61\times10^{57}$. Corchero
\cite{springerlink:10.1023/A:1002714819401} used a similar proof for a
quantum model of neutron stars and we adapt this here for our Skyrme
crystal model:

{\it If there is a locally isotropic, stable solution to
the generalised TOV equation \eqref{TOV} with mass $M$ and total
baryon number $N$, then all locally anisotropic solutions that have
the same total baryon number $N$, in the neighbourhood of that stable
solution, will have a mass not smaller than $M$. }

To prove this,  we first note that stable solutions for a given baryon
number at zero temperature, by definition, have a mass that is not
greater than that which could be achieved by any variation of the
density that preserves the baryon number \cite{Weinberg:100595}.\par

We then consider two changes to a stable solution with mass $M_1$,
baryon number $N_1$, density $\rho_1(r)$ and number density
$n_1(r)$. The first involves changing the density from $\rho_1(r)$ to
$\rho_2(r)$ while keeping the total baryon number constant and
preserving locally isotropy. This will result in a configuration that
has a mass $M_2$ that is greater than or equal to the mass $M_1$ of
our initial configuration as that was defined as the minimum mass
solution. This new configuration will have a different number density,
$n_2(r)$, but the same total baryon number, $N_1$, by assumption.\par 

The second change involves introducing local anisotropy while the
mass, $M_2$, remains the same, as does the density, $\rho_2(r)$. In
order to keep $\rho_2(r)$ constant when we alter the configuration so
that it is now made of anisotropic Skyrme crystal the number density
must also be altered. A change from an isotropic to an anisotropic
Skyrme crystal involves increasing its energy, and therefore mass, so
to keep its mass density constant we need to reduce the number of
Skyrmions to the new number density $n_3(r)\leq n_2(r)$, meaning the
total baryon number is now $N_3\leq N_1$.\par

The two changes described have the effect of firstly increasing $M$
without changing $N$ and then, secondly, decreasing $N$ without
altering $M$. We know that $M$ is a monotonically increasing function
of $N$ for isotropic Skyrme crystal stable star configurations, 
so we have proved
that moving from isotropic to anisotropic Skyrme crystal
configurations increases the energy for a given baryon number so does
not produce a minimum energy solution.

\begin{figure}[!ht]
\begin{center}
\includegraphics[height=6cm,width=12cm, angle=0]{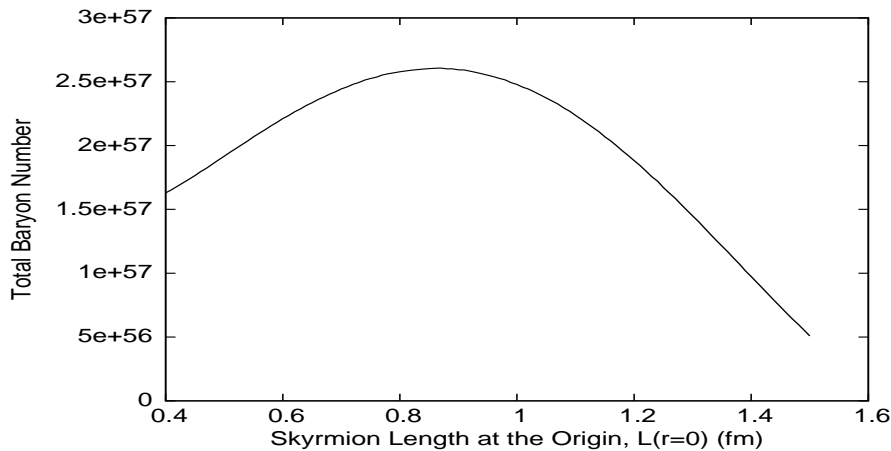}
\caption{\emph{Total baryon number as a function of the size of the 
Skyrmions at the centre of the star, $L(r=0)$.}}
\label{Iso}
\end{center}
\end{figure}

The above proof however does not rule out the existence of anisotropic
Skyrme crystal solutions for those baryon numbers for which there does
not exist an isotropic Skyrme crystal solution and such configurations
will be discussed in the next section.\par

To confirm the results obtained for isotropically deformed crystals, we
will now determine the properties of these symmetric stars by imposing that 
symmetry, {\it i.e.} $p_t=p_r$. In this case the problem simplifies greatly 
and the TOV equation (\ref{eq_mainTOV}) reduces to 
\begin{equation}
 \frac{dp_{r}}{dr}  = -(\rho + p_{r})\frac{m+4\pi r^3 p_r}{r(r-2m)}
\label{eq_mainSymmetricTOV}
\end{equation}
Using this standard TOV equation, a central Skyrmion length
$\lambda_{t}(r=0)=\lambda_{r}(r=0)=L(r=0)$ can be specified at the centre of
the star. The equation can then be numerically integrated over the
radius of the star using the Skyrmion energy equation \eqref{Energy}
with
\begin{equation}
 p_r=-\frac{\partial E}{\partial \lambda_r^3},
\end{equation}
where,  as we are only considering isotropic Skyrme crystal deformations, 
$\lambda_{t}=\lambda_{r}$ and $p=0$ in the energy equation. This was done
using a fourth order Runge Kutta method over points every
$20$m. Notice that this did not require the explicit minimisation of $M_G$.
Figure \ref{Iso} shows a plot of the total baryon number of the
star against its Skyrmion length at the center, $L(r=0)$, 
calculated using this method. \par

We found that isotropic Skyrme crystal solutions can be found up to a
baryon number of $2.61\times10^{57}$, which is equivalent to a mass of
$1.49M_\odot$. This agrees with the results that we found from
our minimisation procedure using the generalised TOV equation that
allows for anisotropic Skyrme crystal deformations. 

\begin{table}
\begin{center}
\begin{tabular}{|c|c|c|c|c|c|}
\hline
$B$ & Total Energy\,\,(J) & Energy/$B$\,\, (J) & Mass/$M_\odot$ & $R({\rm m})$ & $S_{min}$ \\
\hline
$1.0\times 10^{55}$ & $1.16210\times 10^{45}$ & $1.16210\times 10^{-10}$ & 0.00649160 & $2219.20$ & $0.991503$ \\
\hline
$1.0\times 10^{56}$ & $1.15114\times 10^{46}$ & $1.15114\times 10^{-10}$ & 0.0643083 & $4714.35$ & $0.959976$ \\
\hline
$2.0\times 10^{56}$ & $2.28551\times 10^{46}$ & $1.14276\times 10^{-10}$ & 0.127680 & $5875.04$ & $0.936375$ \\
\hline
$4.0\times 10^{56}$ & $4.51669\times 10^{46}$ & $1.12917\times 10^{-10}$ & 0.252325 & $7266.13$ & $0.897929$ \\
\hline
$6.0\times 10^{56}$ & $6.70497\times 10^{46}$ & $1.11750\times 10^{-10}$ & 0.374573 & $8177.42$ & $0.865580$ \\
\hline
$8.0\times 10^{56}$ & $8.85463\times 10^{46}$ & $1.10683\times 10^{-10}$ & 0.494664 & $8852.67$ & 0.835232 \\
\hline
$1.0\times 10^{57}$ & $1.09679\times 10^{47}$ & $1.09679\times 10^{-10}$ & 0.612721 & $9379.47$ & 0.808115 \\
\hline
$1.2\times 10^{57}$ & $1.30461\times 10^{47}$ & $1.08718\times 10^{-10}$ & 0.728823 & $9798.86$ & 0.781969  \\
\hline
$1.4\times 10^{57}$ & $1.50899\times 10^{47}$ & $1.07785\times 10^{-10}$ & 0.842997 & $10133.2$ & 0.755523  \\
\hline
$1.6\times 10^{57}$ & $1.70994\times 10^{47}$ & $1.06871\times 10^{-10}$ & 0.955258 & $10394.6$ & 0.730148 \\
\hline
$1.8\times 10^{57}$ & $1.90741\times 10^{47}$ & $1.05967\times 10^{-10}$ & 1.065578 & 10588.7 & 0.704181 \\
\hline
$2.0\times 10^{57}$ & $2.10132\times 10^{47}$ & $1.05066\times 10^{-10}$ & 1.173903 & 10714.6 & 0.677181 \\
\hline
$2.2\times 10^{57}$ & $2.29147\times 10^{47}$ & $1.04158\times 10^{-10}$ & 1.280129 & 10761.8 & 0.649383 \\
\hline
$2.4\times 10^{57}$ & $2.47750\times 10^{47}$ & $1.032293\times 10^{-10}$ & 1.38406 & 10694.5 & 0.619124 \\
\hline
$2.6\times 10^{57}$ & $2.65860\times 10^{47}$ & $1.022536\times 10^{-10}$ & 1.48522 & 10367.5 & 0.577658 \\
\hline
 
\end{tabular}
\caption{\emph{Properties of the isotropic minimum energy neutron star 
configurations for various baryon numbers.}}
\label{IsoT}
\end{center}
\end{table}

Table \ref{IsoT} shows some of the properties of the minimum energy
solutions for various baryon numbers obtained from the energy
minimisation of the generalised TOV equation. The results are in perfect 
agreement with the results obtained by solving the isotropic TOV equation 
(\ref{eq_mainSymmetricTOV}). The quantity
$S_{min}$ is the minimum value, over the radius of the star, of 
\begin{equation} 
  S(r)=e^{-\lambda(r)}=1-\frac{2m(r)}{r},
\end{equation} 
which appears in the static, spherically symmetric
metric \eqref{metric} that we are considering. The zeros of $S(r)$ correspond 
to singularities in the metric, or in other words, to horizons. Had
$S_{min}$  been negative, we would have concluded that the neutron star 
would have collapsed into a black hole, but this never occurred.

We note that the solutions are energetically favourable as the energy
per baryon decreases when the total baryon number increases,
indicating that the solutions are stable. They correspond to the
solutions to the right of the maximum in figure \ref{Iso} with
solutions to the left being unstable with a higher energy per baryon
for a given baryon number, and therefore not found by the energy
minimisation procedure. \par 

\begin{figure}[!ht]
\begin{center}
\includegraphics[height=6cm,width=12cm, angle=0]{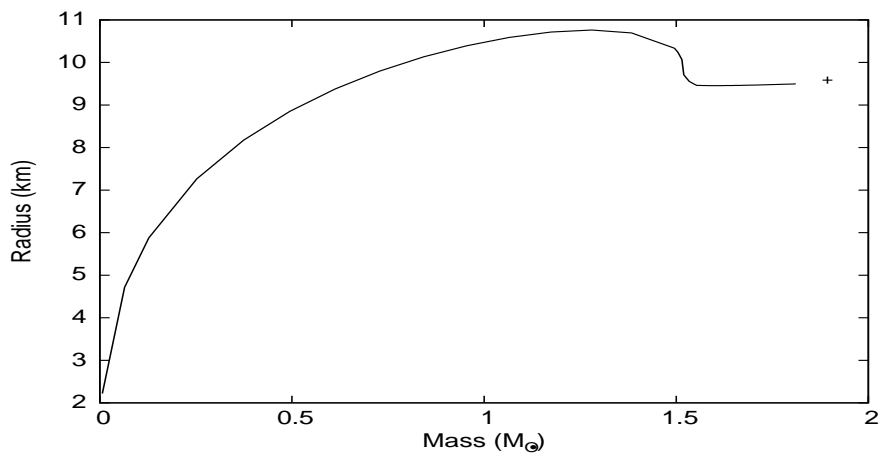}
\caption{\emph{Radius of the neutron star solutions as a function of 
their mass (solid line), and that of the maximum mass solution (cross).
}}
\label{massradius}
\end{center}
\end{figure}

The neutron star solutions which have masses larger than the mass of the Sun
have radii of about $10$km, which very much matches the 
experimental estimates of the radii of observed neutrons stars.
Notice also that the largest neutron star, in our model, has a mass of
approximately
$1.28M_\odot$, and above that value, the radius of the stars decreases 
with their mass (see table \ref{IsoT} and figure \ref{massradius}).

\begin{figure}[!ht]
\begin{center}
\includegraphics[height=6cm,width=12cm, angle=0]{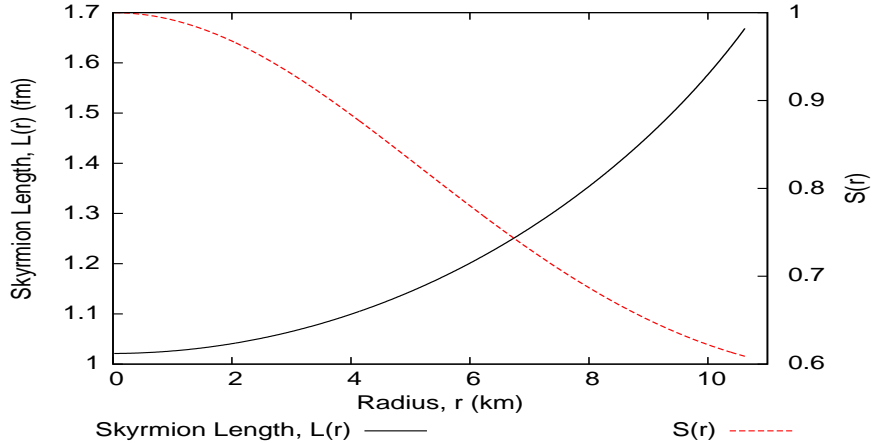}
\caption{\emph{Variation of the size of the isotropic Skyrmions, $L(r)$, 
(solid line) and of the metric function $S(r)$ (dotted line) over the 
radius of a star of mass $1.40M_\odot$.}}
\label{IsoLengths}
\end{center}
\end{figure}

We now consider the structures of these isotropic Skyrme crystal
stars, in particular we consider the case of a star with a mass
of $1.40M_\odot$, a typical mass for a realistic neutron star,
equivalent to a baryon number of $2.44\times 10^{57}$, although all
the isotropic Skyrme crystal minimum energy solutions show the same
qualitative behaviour. Figure \ref{IsoLengths} shows the size of the
Skyrmions, $L(r)$, over the radius of the star. As expected the Skyrmions are
deformed more towards the centre of the star than at the edge,
increasing the Skyrmion mass density by a factor of
$4.44$. Due to this decrease in the size of the Skyrmions as we reach
the centre of the star the stress is higher at the centre and
decreases towards zero at the edge of the star as imposed by
the boundary conditions. \par

The isotropic Skyrme crystal solutions have a $S_{min}$ that
is always greater than zero so the configurations do not
collapse into black holes. Figure \ref{IsoLengths} also shows how the value 
of $S(r)$ varies over the radius of the star.\par

\subsection{Stars Made of Anisotropically Deformed Skyrme Crystal}
Having shown in the previous section that no isotropic Skyrme crystal 
solutions exist for baryon numbers larger than $2.61\times10^{57}$, we will 
now show that anisotropic solutions do exist.

Table \ref{AnIsoT} shows some of the
properties of the anisotropic minimum energy Skyrme crystal solutions
for various baryon numbers obtained using the generalised TOV
equation. We found solutions in this way up to a baryon number of
$3.25\times10^{57}$, corresponding to $1.81M_\odot$, after which the 
numerical energy minimisation procedure became difficult to implement. However
by using a similar simulated annealing process to maximise the baryon
number, rather than minimise the energy for a particular baryon
number, we found anisotropic Skyrme crystal solutions up to a baryon
number of $3.41\times10^{57}$, equivalent to $1.90M_\odot$. At
this maximum baryon number solution there is only one possible
configuration of the Skyrmions, as any modification to it results
in a decrease in the baryon number, hence it is the minimum energy
solution. Above this baryon number, solutions do not exist.

\begin{table}
\begin{center}
\begin{tabular}{|c|c|c|c|c|c|}
\hline
$B$ & Total Energy\,\,(J) & Energy/$B$\,\, (J) & Mass/$M_\odot$ & $R({\rm m})$ & $S_{min}$ \\
\hline
$2.65\times 10^{57}$ & $2.70277\times 10^{47}$ & $1.01991\times 10^{-10}$ & 1.50990 & 10091.8 & 0.559060 \\
\hline
$2.70\times 10^{57}$ & $2.74605\times 10^{47}$ & $1.01706\times 10^{-10}$ & 1.53408 & 9555.51 & 0.526465 \\
\hline
$2.75\times 10^{57}$ & $2.78943\times 10^{47}$ & $1.01434\times 10^{-10}$ & 1.55832 & 9460.46 & 0.514207 \\
\hline
$2.80\times 10^{57}$ & $2.83310\times 10^{47}$ & $1.01182\times 10^{-10}$ & 1.58271 & 9456.89 & 0.506402 \\
\hline
$2.85\times 10^{57}$ & $2.87706\times 10^{47}$ & $1.00949\times 10^{-10}$ & 1.60727 & 9456.46 & 0.498735 \\
\hline
$2.90\times 10^{57}$ & $2.92133\times 10^{47}$ & $1.00735\times 10^{-10}$ & 1.63200 & 9457.92 & 0.491152 \\
\hline
$2.95\times 10^{57}$ & $2.96592\times 10^{47}$ & $1.00540\times 10^{-10}$ & 1.65691 & 9460.65 & 0.483633 \\
\hline
$3.00\times 10^{57}$ & $3.01087\times 10^{47}$ & $1.00362\times 10^{-10}$ & 1.68202 & 9465.06 & 0.476231  \\
\hline
$3.05\times 10^{57}$ & $3.05619\times 10^{47}$ & $1.00203\times 10^{-10}$ & 1.70734 & 9469.97 & 0.468880  \\
\hline
$3.10\times 10^{57}$ & $3.10191\times 10^{47}$ & $1.00062\times 10^{-10}$ & 1.73288 & 9475.76 & 0.461631  \\
\hline
$3.15\times 10^{57}$ & $3.14807\times 10^{47}$ & $9.99388\times 10^{-11}$ & 1.75867 & 9481.95 & 0.454438  \\
\hline
$3.20\times 10^{57}$ & $3.19472\times 10^{47}$ & $9.98351\times 10^{-11}$ & 1.78473 & 9489.04 & 0.447382  \\
\hline
$3.25\times 10^{57}$ & $3.24191\times 10^{47}$ & $9.97510\times 10^{-11}$ & 1.81109 & 9496.62 & 0.440435  \\
\hline
 
\end{tabular}
\caption{\emph{Properties of the anisotropic minimum energy neutron star 
configurations for various baryon numbers.}}
\label{AnIsoT}
\end{center}
\end{table}

As in the case of isotropic Skyrme crystal deformations 
we find that the solutions are
energetically favourable as the energy per baryon decreases as the
total baryon number increases, indicating stable solutions. 
As the baryon number is
increased towards its maximum value of $3.41\times10^{57}$ the energy
per baryon begins to level off and we find that the maximum baryon
number has the lowest energy per baryon, as in the isotropic
case. \par 

\begin{figure}[!ht]
\begin{center}
\includegraphics[height=6cm,width=12cm, angle=0]{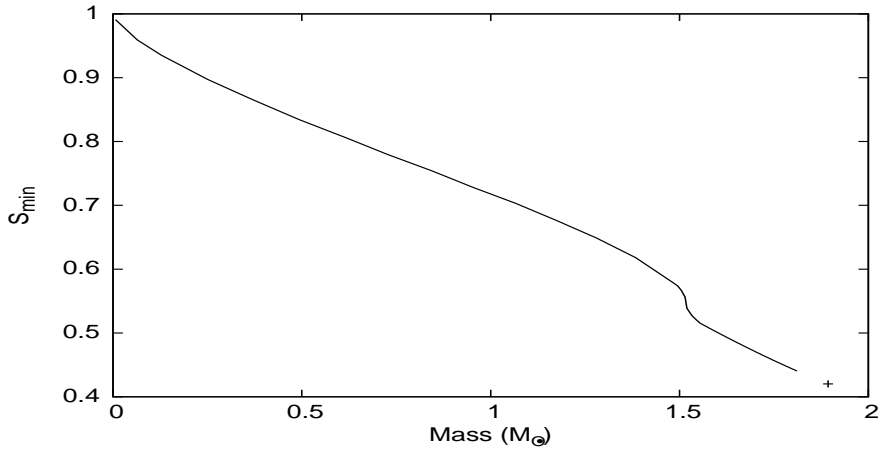}
\caption{\emph{$S_{min}$ of the neutron star solutions as a function of 
their mass. The maximum mass solution is shown as a cross.}}
\label{Sminmass}
\end{center}
\end{figure}

We can see that the configurations we have constructed do not
collapse into a black hole by noticing that the values of $S_{min}$ are
always positive, as shown in figure \ref{Sminmass}. 

Figure \ref{massradius} shows a plot of the mass radius
curve for both the isotropic and anisotropic Skyrme crystal cases,
with the mass in units of $M_\odot$. 
As stated above, large isotropic crystal neutron stars have a radius
that decreases as the mass increases. We can clearly see in figure 
\ref{massradius},
that at the critical mass of $1.49M_\odot$, the radius keeps decreasing 
as the mass of the star increases. Moreover, we also observe a sharp drop of 
radius just over $1.5M_\odot$ followed by a plateau at about $9.5$km.
  
\begin{figure}
  \centering  
\subfloat[]
{\label{L_LARGE}\includegraphics[height=4cm,width=7cm, angle=0]{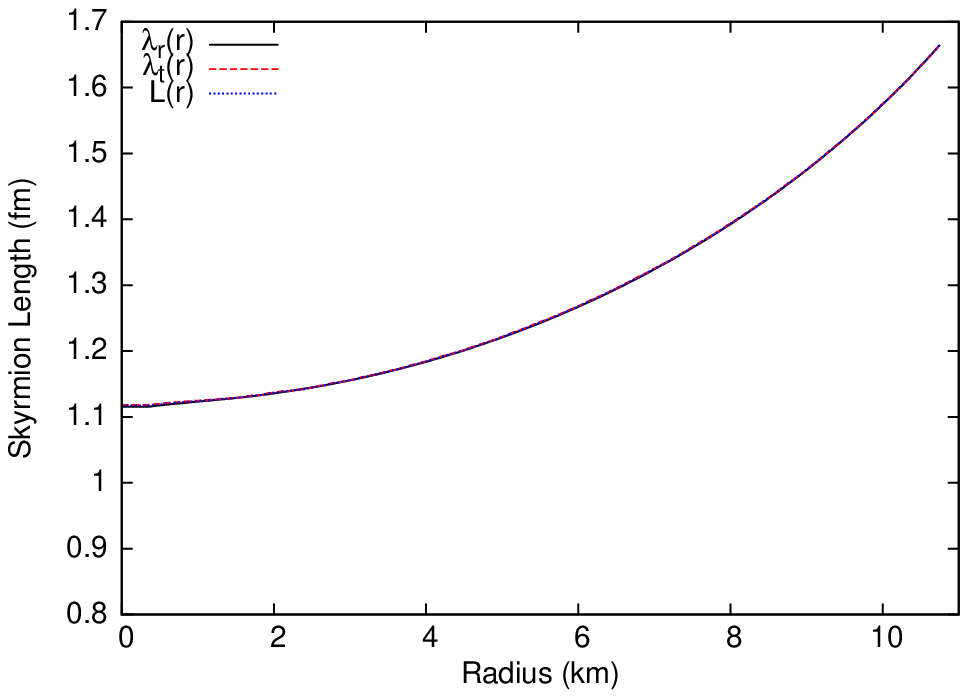}} 
\subfloat[]
{\label{L_ISOHEAVY}\includegraphics[height=4cm,width=7cm, angle=0]{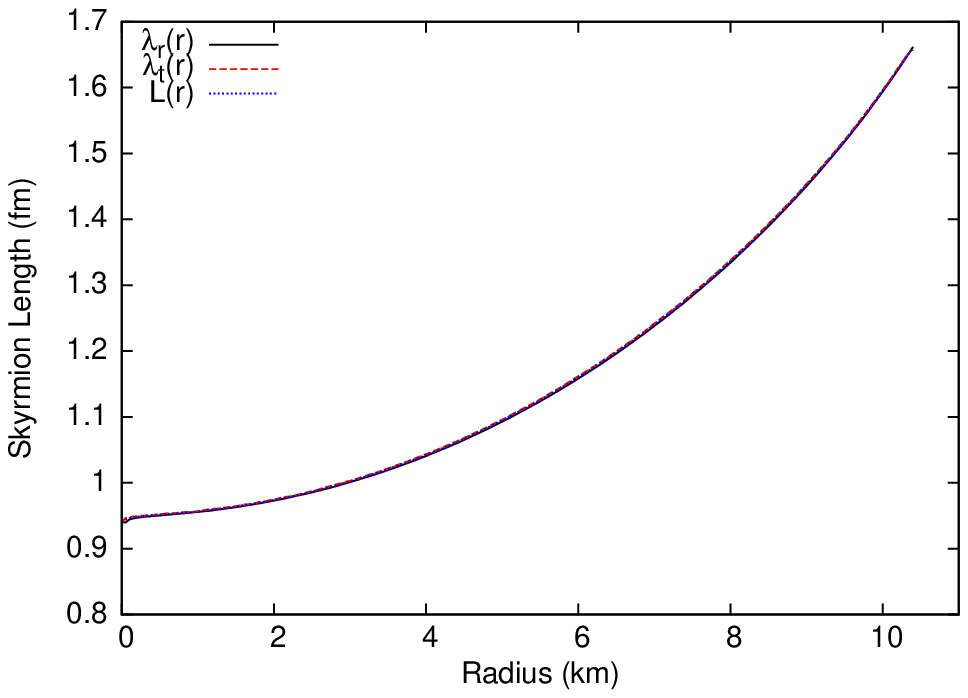}}\\
\subfloat[]
{\label{L_DENS}\includegraphics[height=4cm,width=7cm, angle=0]
{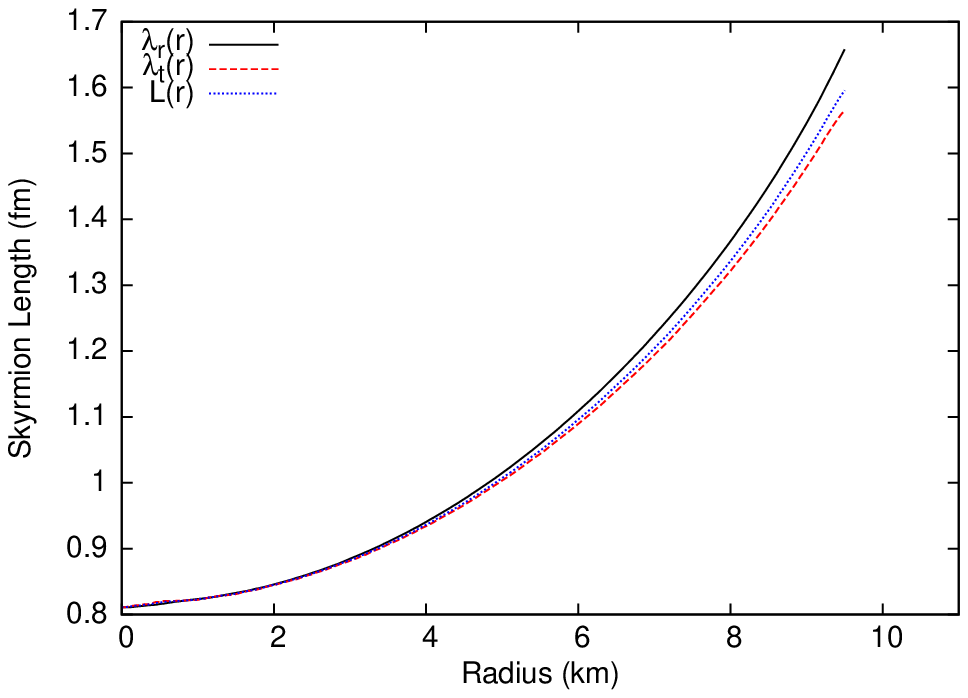}} 
\subfloat[]
{\label{L_MAX}\includegraphics[height=4cm,width=7cm, angle=0]
{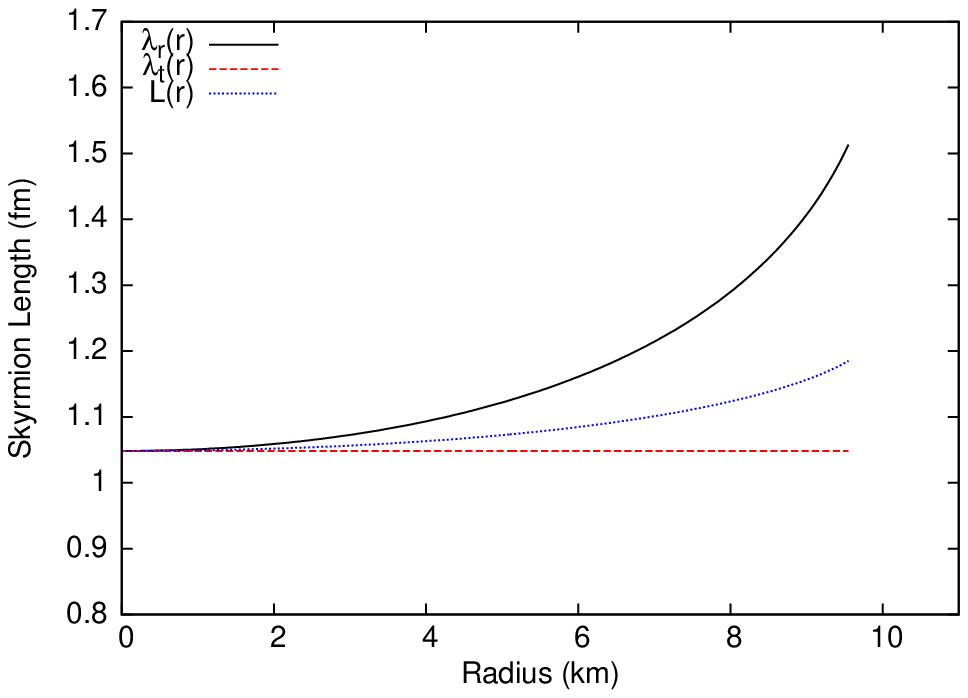}} 
 \caption{
Skyrmion lengths $\lambda_r(r)$ (solid line), $\lambda_t(r)$ (dashed line)
and $L(r)$ (dotted line) for 
a) Largest neutron star ($R=10.8$km):  $M=1.28M_\odot$ 
b) Heaviest isotropic neutron star:  $M=1.49M_\odot$ 
 (all lengths coincide as they are made of isotropically deformed crystal); 
c) Densest neutron star: $M=1.54M_\odot$;
d) Heaviest neutron star: $M=1.90M_\odot$.}
\label{AnIso}
\end{figure}

By considering anisotropic as well as isotropic Skyrme crystal
solutions we have extended the mass range over which solutions can be
found, finding masses up to $28\%$ above the maximum mass of the
isotropic case. This is an interesting finding because isotropy of
matter is often taken as an assumption when studying neutron star
models, including the Skyrme crystal case considered in
\cite{Walhout:1987ce},\cite{Walhout:1990va},\cite{Jaikumar:2005gq}
and a maximum mass is
then derived. We have shown that by not assuming isotropy and instead
allowing anisotropic matter configurations the maximum mass can be
increased by a significant amount. In this simple Skyrme crystal model
the maximum mass found is equivalent to $1.90M_\odot$ and the
recent discovery of a $1.97 \pm 0.04\,M_\odot$ neutron star
\cite{Demorest:2010bx}, the highest neutron star mass ever determined,
makes this an encouraging finding, especially when we consider that
including the effects of rotation into our model will increase the
maximum mass found, by up to $2\%$ for a star with a typical $3.15$ms
spin period \cite{Berti:2004ny}. \par

Figure \ref{AnIso} shows a selection of plots of the Skyrmion lengths
$\lambda_r$ and $\lambda_t$ and the Skyrmion size
$L$, equation (\ref{eq_L}), over the radius of the star for four special 
stars:  
the largest star, with radius 
$R=10.8$km and mass $M=1.28M_\odot$ (figure \ref{L_LARGE}); 
the heaviest isotropically deformed star $M=1.49M_\odot$ 
(figure \ref{L_ISOHEAVY}); 
the densest neutron star, $M=1.54M_\odot$ (figure \ref{L_DENS}) and
the heaviest neutron star, $M=1.90M_\odot$ (figure \ref{L_MAX}). 
The first two are made out of an
isotropically deformed crystal, while the last two are anisotropically 
deformed and one notices that the 
amount of anisotropy increases as the mass increases (the divergence between
$\lambda_r$ and $\lambda_t$ increases).
Throughout this paper, we will use these four special stars 
as examples to illustrate various properties of the neutron stars.

As the maximum
mass is approached the gradient of the profile of tangential
Skyrmion lengths over the radius of the star becomes smaller and we
note that physically meaningful stars composed of anisotropically deformed
crystal
should have $d\lambda_t/dr\geq0$ \cite{2003}. This confirms that the
minimum energy solution for the maximum mass found,
$1.90M_\odot$, for anisotropic Skyrme crystal solutions is the
configuration with a constant tangential Skyrmion length as illustrated in 
figure \ref{L_MAX}.

The generalised TOV equation imposes that the
sizes of the Skyrmions are equal in all directions at the centre of
the star, but away from the centre, for all the anisotropic Skyrme
crystal solutions, we find that the amount of Skyrmion anisotropy
increases as we move towards the edge of the star, reaching the maximum at
the edge. The Skyrmions are deformed to a greater extent in the
tangential direction in agreement with the value of the aspect ratio,
$p$, being negative over the values where $\lambda_r \neq \lambda_t$. 

\begin{figure}
  \centering
\includegraphics[height=6cm,width=12cm, angle=0]{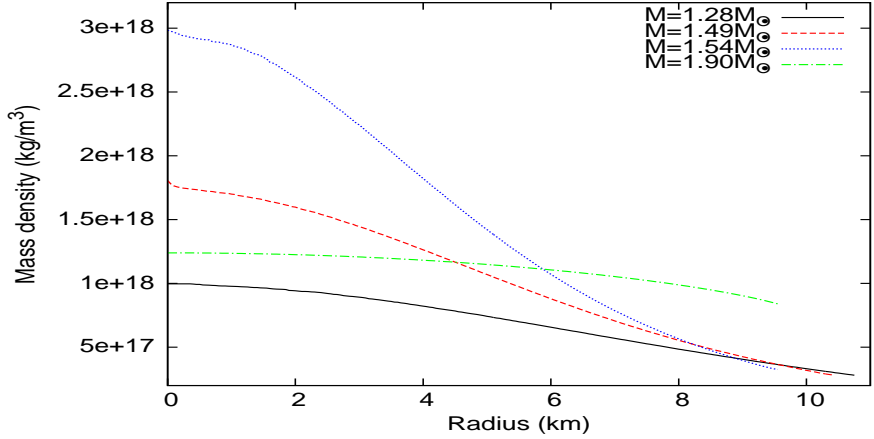}
  \caption{Mass density $\rho(r)$ for: 
a) Largest neutron star ($R=10.8$km):  $M=1.28M_\odot$ (solid line)
b) Heaviest isotropic neutron star:  $M=1.49M_\odot$ (dashed line);
c) Densest neutron star: $M=1.54M_\odot$ (dotted line); 
d) Heaviest neutron star: $M=1.90M_\odot$ (dash dotted line).}
  \label{Massdensity}
\end{figure}

As expected, the profiles for $\lambda_r$ and $\lambda_t$ show that
the mass density at the centre of the star is higher than at the edge,
decreasing monotonically as the radial distance increases. This is
shown by figure \ref{Massdensity} for the largest, heaviest isotropic,
densest and maximum mass solutions. 

\begin{figure}[!ht]
\begin{center}
\includegraphics[height=6cm,width=12cm, angle=0]{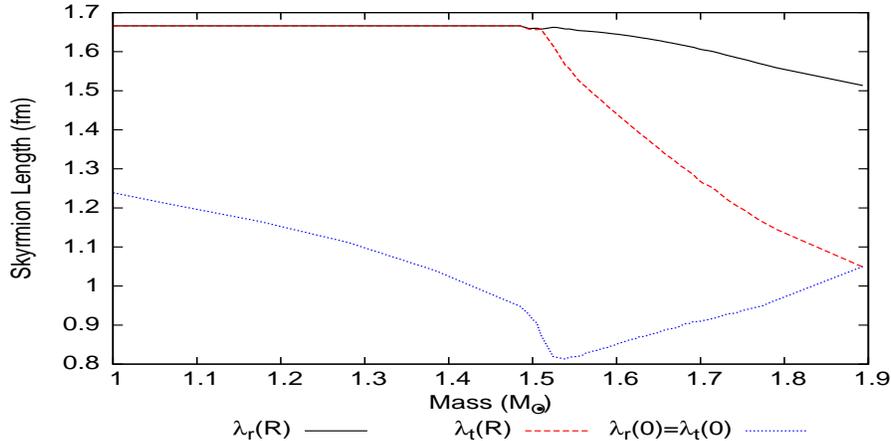}
\caption{\emph{Skyrmion lengths at the edge of the star, 
$\lambda_r(R)$ (solid line)
and $\lambda_t(R)$ (dashed line), and at the center of the star, 
$\lambda_r(0)=\lambda_t(0)$ (dotted line), as a function of the star mass.}}
\label{Lengths}
\end{center}
\end{figure}

In figure \ref{Lengths} one can see how the lengths of the Skyrme crystal
$\lambda_r$ and $\lambda_t$ vary with the mass of the star both at the center 
($r=0$) and the edge of the star ($r=R$). For isotropically deformed stars,
$\lambda_r(R)=\lambda_t(R)$ is constant and corresponds to the minimum energy
Skyrme crystal in the absence of gravity. Not surprisingly, 
$\lambda_r(0)=\lambda_t(0)$ decreases steadily as the mass of the star 
increases, showing that the density at the center of the star increases.
Once the phase transition has taken place and the star is too heavy
to remain isotropically deformed, we observe that $\lambda_r(0)=\lambda_t(0)$
drops sharply to a local minimum, reached for $M\approx 1.54M_\odot$.
Meanwhile, $\lambda_r(R)$ and $\lambda_t(R)$ remain nearly identical.
Beyond the minimum of $\lambda_{r,t}(0)$, 
$\lambda_r(R)$ and $\lambda_t(R)$ start to diverge sharply; $\lambda_r(R)$
decreases slightly in value while $\lambda_t(R)$ decreases rapidly. These
stars are thus much more compressed in the tangential direction than in the 
radial one. As seen on figure \ref{L_MAX}, $\lambda_t(R)=\lambda_t(0)$ 
for the maximum mass neutron star.

\begin{figure}
  \centering 
\includegraphics[height=6cm,width=12cm, angle=0]{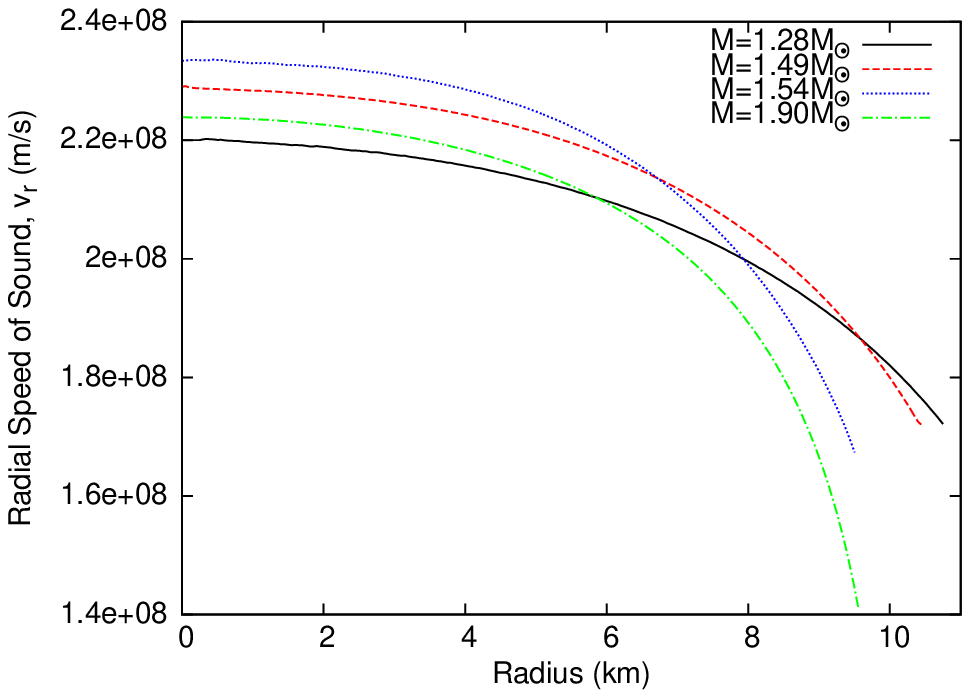}
  \caption{Radial speed of sound, $v_r(r)$ for 
a) Largest neutron star ($R=10.8$km):  $M=1.28M_\odot$ (solid line)
b) Heaviest isotropic neutron star:  $M=1.49M_\odot$ (dashed line); 
c) Densest neutron star: $M=1.54M_\odot$ (dotted line); 
d) Heaviest neutron star: $M=1.90M_\odot$ (dash dotted line).} 
  \label{sound}
\end{figure}

Another property of a neutron star worth considering is the speed of sound.
To compute it one needs to know how the 
energy of the crystal varies when it is deformed in the direction of 
wave propagation. Using (\ref{Energy}) we can thus compute the speed of sound
in the $z$ direction. 
To compute the speed of sound in the $x$ and $y$ directions when the crystal 
is deformed we need to know how the energy of the crystal varies when 
the crystal is deformed in all three directions independently, an expression
we do not have. 

We are thus only able to compute the radial speed of sound inside a neutron star
and it is given by
\begin{equation}
v_r = \left(\frac{d p_r}{d \lambda_r}
         \left(\frac {d \rho}{d\lambda_r} \right)^{-1}\right)^{1/2}
\end{equation}
where both $p_r$ and $\rho$ are functions of $\lambda_r$ and $\lambda_t$ 
given respectively by (\ref{EOS}) and (\ref{eq_rho}).
Obviously, when the crystal inside the star is isotropically deformed, 
the speed of sound is the same in all 3 directions.

First of all it is interesting to notice that the speed of sound in the
minimum energy
Skyrme crystal, in the absence of a gravitational field, is amazingly large:
$v=0.57\,c$. This is the speed of sound at the surface of a neutron star
when it is deformed isotropically. From figure \ref{sound} one sees that 
$v_r$ increases as one moves towards the center of the star.
As $v_r$ is directly related to the density of the star, it
is not surprising to find that
the maximum radial speed, $v_r= 0.78c$, is reached at the center
of the densest neutron star, {\it i.e.} the one with $M=1.54M_\odot$.
As expected, $v_r < c$ everywhere.

\begin{figure}
  \centering 
\includegraphics[height=6cm,width=12cm, angle=0]{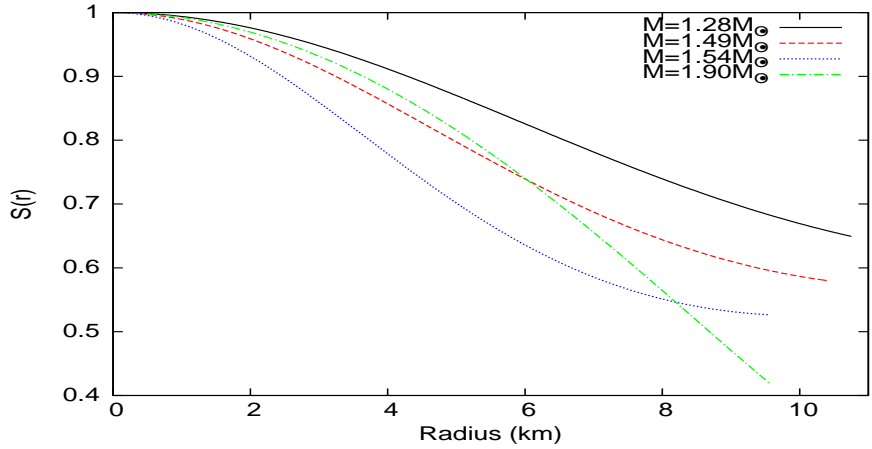}
  \caption{The function $S(r)$ for: 
a) Largest neutron star ($R=10.8$km):  $M=1.28M_\odot$ (solid line)
b) Heaviest isotropic neutron star:  $M=1.49M_\odot$ (dashed line);
c) Densest neutron star: $M=1.54M_\odot$ (dotted line); 
d) Heaviest neutron star: $M=1.90M_\odot$ (dash dotted line).} 
  \label{S}
\end{figure}

Figure \ref{S} shows how the value of $S(r)$ varies over the radius of
the star for, again, the largest, heaviest isotropic, densest and maximum 
mass solutions, showing how the
metric is altered as $r$ varies. The minimum value of
$S(r)$ is always located at the edge of the star, {\it i.e.} $S_{min}=S(R)$,
and it is presented in figure \ref{Sminmass} as a function of
the star masses. One sees that $S_{min}$ decreases monotonically
as the mass increases, and exhibits a sharp decrease just over $1.5M_\odot$, 
{\it i.e.} just above the critical mass.
However $S_{min}$ always remains positive, indicating that no black hole 
is formed.\par

\begin{figure}[!ht]
\begin{center}
\includegraphics[height=6cm,width=12cm, angle=0]{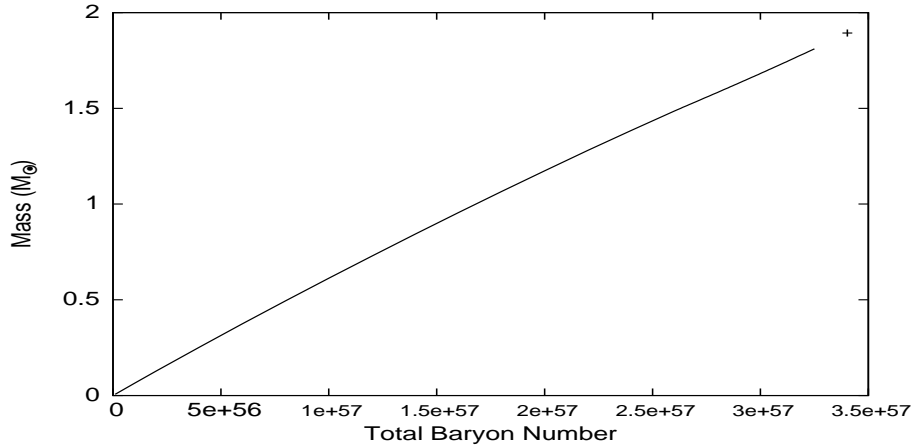}
\caption{\emph{Mass of the neutron star solutions as a function of 
their baryon number. The maximum mass solution is shown as a cross.}}
\label{BM}
\end{center}
\end{figure}

Figure \ref{BM} shows how the total baryon number and the mass of all
the solutions found are related. As the baryon number increases the
effects of gravitational attraction increase, resulting in a slightly lower
gravitational mass per baryon than expected from a linear relation.

We note that the minimum value of the aspect ratio, $p$, for the
minimum energy configurations found is $-0.283$ and the minimum value
of $L$ is $8.11\times10^{-16}$, both of which are within the valid range of
values for equation \eqref{Energy} \cite{Castillejo1989801}.\par

\subsection{Inclusion of the Pion Mass}

Throughout the work described we have assumed a zero pion mass. The inclusion 
of a non-zero pion mass can be considered by including the pion mass term, 
\begin{equation}
 \int \frac{m_{\pi}^2F_{\pi}^2}{8}{\rm{Tr}}(U-1)d^3x,
\end{equation}
in the static Skyrme Lagrangian \eqref{Lagrangian}, where $U$ is the Skyrme 
field, $F_\pi$ is the pion decay constant and $m_\pi$ is the pion mass. 
Using the cubic lattice of $\alpha$-like Skyrmions that has been 
considered above one finds that 
${\rm{Tr}}(U-1)=-2$, meaning that the energy $E_\pi$ arising from the pion 
mass term reduces to
\begin{equation}
 E_\pi=\frac{1}{4} m_{\pi}^2F_{\pi}^2L^3,
\label{eq_Epi}
\end{equation}
an energy term proportional to the volume of the Skyrmions. \par

\begin{figure}[!ht]
\begin{center}
\includegraphics[height=6cm,width=12cm, angle=0]{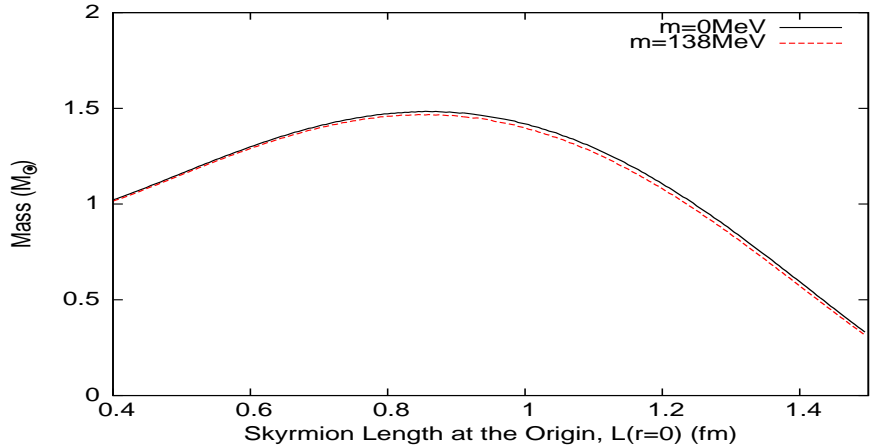}
\caption{\emph{Mass of the star as a function of the size of the Skyrmions 
at the centre, $L_0$, for zero pion mass (solid line) and $m=138$MeV 
(dashed line).}}
\label{pionmass}
\end{center}
\end{figure}

It can be seen in figure \ref{pionmass} that including a pion mass of 
$m=138$MeV decreases the
maximum mass of the star by a very small amount from $1.49$ to $1.47M_\odot$ 
while also slightly decreasing the central density at which
this occurs.\par

Including a pion mass of $m=138$MeV in the simulated annealing process
used to find the maximum baryon number for the anisotropic Skyrme
crystal solutions results in a maximum baryon number of
$3.34\times10^{57}$, equivalent to $1.88M_\odot$, a decrease of
$0.02M_\odot$ from the maximum mass found in the case without a
pion mass.\par

This gives an indication as to how the pion mass affects the
structures of the neutron star configurations that can be constructed,
and a similar reduction in the maximum mass is expected for all the
anisotropic crystal solutions, however when the pion mass is included
it also has the effect of driving the Skyrme crystal lattice away from
the half-Skyrmion symmetry \cite{Castillejo1989801}. This will be a
small effect for the dense Skyrme crystals that we are considering
because while the pion mass term is the dominant term in the
Lagrangian far away from the centres of the Skyrmions when they are
well separated, in the dense Skyrme crystal there is no space away
from the centres of the Skyrmions so it becomes less important in
affecting the field distributions. Its effect will be to reduce the
pion mass term, Eq. (\ref{eq_Epi}), by a small amount.\par

\subsection{Stars above the Maximum Mass}
As in other studies of neutron starts based on the Skyrme model, we found 
a critical mass above which solutions do not exist. In other words,
when the star is too  massive, the crystal of which it is made is not 
capable of counterbalancing the gravitation pull and the star then 
collapses into a black hole. This is indeed what we observed when trying to
construct solutions above the critical mass: the energy of the configuration 
kept decreasing as the radius of the star decreased and the $S_{min}$ function
became negative, indicating the formation of an horizon, and 
hence a black hole.

Throughout this work we have assumed a spherically symmetric metric
and stress tensor, however, these assumptions could be removed and it
may be that higher mass solutions could be found. We could instead
consider an axially symmetric metric, the most general form
\cite{PhysRev.153.1359} being 
\begin{equation}
 ds^2=\alpha^2(d\rho^2+dz^2)+\beta^2d\phi^2-\gamma^2dt^2,
\end{equation}
when written in cylindrical coordinates. The stress tensor,
\begin{equation}
 T_\nu^\mu={\rm diag}(\rho,p_1,p_2,p_3),
\end{equation} 
could then be completely anisotropic with $p_1\neq
p_2\neq p_3$. Minimum energy solutions to Einstein's equations for
such a metric and stress tensor could be found by direct minimisation
of the action of the Skyrme model coupled to gravity or by using an,
as yet undetermined, axisymmetric form of the TOV equation. Another
approach to investigate such solutions would be to perturb the
spherically symmetric solutions that we have found. Following the
procedure for doing so described in \cite{PhysRev.153.1359} the
exterior metric for an axially symmetric solution can be written in
Schwarzschild coordinates and, after comparing the exterior
spherically symmetric Schwarzschild solution to our solutions for the
interior metric of the star and finding the substitutions necessary to
move from one to the other, we can make to same substitutions to the
axially symmetric exterior metric. This allows us to then describe
approximately both the metric and the stress energy tensor of the
axially symmetric solution. To carry out such investigations into
axially symmetric static configurations an equation analogous to
\eqref{Energy} which would relate the energy of the Skyrme crystal to
its size and deformation in all three directions independently would
need to be considered.  \par

We have also assumed that the stress tensor, $T_\nu^\mu={\rm
diag}(\rho,p_r,p_\theta,p_\phi)$, is diagonal, however, if shear
strains are included in our model off diagonal components would have
to be introduced. This would also remove the assumption of spherical
symmetry altering the configurations found.\par

Spherical symmetry also needs to be removed to consider rotating
stars. This will result in configurations above the maximum mass found
in this work, by up to $2\%$ for a star with a typical $3.15$ms spin
period \cite{Berti:2004ny}, and as neutron stars are known to be
rotating, this is an important effect to consider.\par

\section{Conclusions}
Neutron stars are large bodies of matter where the electrons, instead of 
circling atoms, are forced to merge with the nuclei, resulting in  
extremely dense stars made entirely of neutrons. Their temperature,
from a nuclear point of view, is very low and this means nuclear matter
must be considered as a solid rather than a fluid. Moreover, the 
gravitational pull of the star is so strong that the 
``atmospheric'' fluid one might expect at the surface is of 
negligible height.

In this context, the Skyrme model, known to be a low energy effective field 
theory for QCD \cite{Witten:1983tw}, is an ideal candidate to describe  
neutron stars once the model is coupled to gravity. The
minimum energy configuration of large numbers of Skyrmions is a cubic 
crystal made of $B=4$ Skyrmions which correspond to a crystal of 
$\alpha$-like particles. We have thus used these solutions as a building block
to describe the neutron star by combining the deformation energy computed in
\cite{Castillejo1989801} and a generalised version of the TOV equation 
\cite{Tolman:1939jz}, \cite{Oppenheimer:1939ne}, \cite{1974ApJ...188..657B}
which describes the static 
equilibrium between matter forces, within a solid or fluid, and the 
gravitational forces self-generated by the matter for a spherically 
symmetric body. 

The key feature of our approach to the problem was to consider the star as a 
solid that could potentially deform itself anisotropically. We then found that 
below $1.49M_\odot$, all stars were made of a crystal deformed 
isotropically, {\it i.e.} the radial strain was identical to the tangential one.
Above that critical value, the neutron star undergoes a critical phase 
transition and the
lattice of Skyrmions compresses anisotropically: the Skyrmions are more 
compressed tangentially than radially. Stars were shown to exist up to a 
critical mass of $1.90M_\odot$, a result that closely matches 
the recent discovery of Demorest et al. \cite{Demorest:2010bx} who measured
the mass of the heaviest neutron star found to date, PSR J1614-2230, 
to be $1.97M_\odot$. We also observed that the maximum radius for a 
Skyrmion star was approximately 11km, a figure that matches well 
the experimental estimations.

In our model we did not consider the rotational energy of the star which is
approximated at about 2\% of its total energy. If we included that extra 
energy, our upper bound would thus 
just fit above the mass of PSR J1614-2230.

Finally we have also shown that if the mass of a 
neutron star was to be raised to cross the critical mass threshold, 
it would collapse into a black hole. 

\section{Acknowledgements}
BP was supported by the STFC Consolidated Grant ST/J000426/1  
and SN by an EPSRC studentship.

\bibliography{mybib}{}

\begin{thebibliography}{10}

\bibitem{Skyrme:1961vq}
T.H.R. Skyrme.
\newblock {A Nonlinear field theory}.
\newblock {\em Proc.Roy.Soc.Lond.}, A260:127--138, 1961.

\bibitem{Skyrme:1962vh}
T.H.R. Skyrme.
\newblock {A Unified Field Theory of Mesons and Baryons}.
\newblock {\em Nucl.Phys.}, 31:556--569, 1962.

\bibitem{Witten:1983tw}
Edward Witten.
\newblock {Global Aspects of Current Algebra}.
\newblock {\em Nucl.Phys.}, B223:422--432, 1983.

\bibitem{Braaten:1990}
W.~Y. Crutchfield, N.~J. Snyderman, and V.~R. Brown.
\newblock Deuteron in the skyrme model.
\newblock {\em Phys. Rev. Lett.}, 68:1660--1662, Mar 1992.

\bibitem{Battye:1997wf}
Richard~A. Battye and Paul~M. Sutcliffe.
\newblock {A Skyrme lattice with hexagonal symmetry}.
\newblock {\em Phys. Lett.}, B416:385--391, 1998.

\bibitem{Battye:2009ad}
Richard~A. Battye, Nicholas~S. Manton, Paul~M. Sutcliffe, and Stephen~W. Wood.
\newblock {Light Nuclei of Even Mass Number in the Skyrme Model}.
\newblock {\em Phys.Rev.}, C80:034323, 2009.

\bibitem{Castillejo1989801}
L.~Castillejo, P.~S.~J. Jones, A.~D. Jackson, J.~J.~M. Verbaarschot, and
  A.~Jackson.
\newblock Dense skyrmion systems.
\newblock {\em Nuclear Physics A}, 501(4):801 -- 812, 1989.

\bibitem{baldo1999nuclear}
M.~Baldo.
\newblock {\em Nuclear methods and the nuclear equation of state}.
\newblock International review of nuclear physics. World Scientific, 1999.

\bibitem{Tilley19921}
D.R. Tilley, H.R. Weller, and G.M. Hale.
\newblock Energy levels of light nuclei a = 4.
\newblock {\em Nuclear Physics A}, 541(1):1 -- 104, 1992.

\bibitem{Barnes:1997}
Chris Barnes, Kim Baskerville, and Neil Turok.
\newblock Normal modes of the
  $\mathit{B}\phantom{\rule{0ex}{0ex}}=\phantom{\rule{0ex}{0ex}}4$ skyrme
  soliton.
\newblock {\em Phys. Rev. Lett.}, 79:367--370, Jul 1997.

\bibitem{Lin:2008}
W.~T. Lin and B.~Piette.
\newblock Skyrmion vibration modes within the rational map ansatz.
\newblock {\em Phys. Rev. D}, 77:125028, Jun 2008.

\bibitem{Walhout:1987ce}
T.S. Walhout.
\newblock {Dense Matter In The Skyrme Model}.
\newblock {\em Nucl.Phys.}, A484:397, 1988.

\bibitem{Tolman:1939jz}
Richard~C. Tolman.
\newblock {Static solutions of Einstein's field equations for spheres of
  fluid}.
\newblock {\em Phys.Rev.}, 55:364--373, 1939.

\bibitem{Oppenheimer:1939ne}
J.R. Oppenheimer and G.M. Volkoff.
\newblock {On Massive neutron cores}.
\newblock {\em Phys.Rev.}, 55:374--381, 1939.

\bibitem{1974ApJ...188..657B}
R.~L. Bowers and E.~P.~T. Liang.
\newblock {Anisotropic Spheres in General Relativity}.
\newblock {\em Astrophys. J.}, 188:657, 1974.

\bibitem{Walhout:1990va}
T.S. Walhout.
\newblock {The Equation of state of dense skyrmion matter}.
\newblock {\em Nucl.Phys.}, A519:816--830, 1990.

\bibitem{Jaikumar:2005gq}
Prashanth Jaikumar and Rachid Ouyed.
\newblock {Skyrmion stars: Astrophysical motivations and implications}.
\newblock {\em Astrophys.J.}, 639:354--362, 2006.

\bibitem{Piette:2007wd}
Bernard~M.A.G. Piette and Gavin~I. Probert.
\newblock {Towards skyrmion stars: Large baryon configurations in the
  Einstein-Skyrme model}.
\newblock {\em Phys.Rev.}, D75:125023, 2007.

\bibitem{Nelmes:2011zz}
Susan Nelmes and Bernard~M.A.G. Piette.
\newblock {Skyrmion stars and the multilayered rational map ansatz}.
\newblock {\em Phys.Rev.}, D84:085017, 2011.

\bibitem{Houghton:1997kg}
Conor~J. Houghton, Nicholas~S. Manton, and Paul~M. Sutcliffe.
\newblock {Rational maps, monopoles and Skyrmions}.
\newblock {\em Nucl.Phys.}, B510:507--537, 1998.

\bibitem{Adkins:1983ya}
Gregory~S. Adkins, Chiara~R. Nappi, and Edward Witten.
\newblock {Static Properties of Nucleons in the Skyrme Model}.
\newblock {\em Nucl.Phys.}, B228:552, 1983.

\bibitem{Kugler1988491}
M.~Kugler and S.~Shtrikman.
\newblock A new skyrmion crystal.
\newblock {\em Physics Letters B}, 208(3-4):491 -- 494, 1988.

\bibitem{springerlink:10.1023/A:1002714819401}
E.S. Corchero.
\newblock Quantum approach to neutron stars leading to configurations with
  local anisotropy and mass above the oppenheimer-volkoff limit.
\newblock {\em Astrophysics and Space Science}, 275:259--274, 2001.
\newblock 10.1023/A:1002714819401.

\bibitem{Weinberg:100595}
Steven Weinberg.
\newblock {\em Gravitation and Cosmology: Principles and Applications of the
  General Theory of Relativity}.
\newblock Wiley, New York, NY, 1972.

\bibitem{Demorest:2010bx}
Paul Demorest, Tim Pennucci, Scott Ransom, Mallory Roberts, and Jason Hessels.
\newblock {Shapiro Delay Measurement of A Two Solar Mass Neutron Star}.
\newblock {\em Nature}, 467:1081--1083, 2010.

\bibitem{Berti:2004ny}
Emanuele Berti, Frances White, Asimina Maniopoulou, and Marco Bruni.
\newblock {Rotating neutron stars: an invariant comparison of approximate and
  numerical spacetime models}.
\newblock {\em Mon. Not. Roy. Astron. Soc.}, 358:923--938, 2005.

\bibitem{2003}
M.~K. Mak and T.~Harko.
\newblock Anisotropic stars in general relativity.
\newblock {\em Proceedings: Mathematical, Physical and Engineering Sciences},
  459(2030):pp. 393--408, 2003.

\bibitem{PhysRev.153.1359}
Walter~C. Hernandez.
\newblock Static, axially symmetric, interior solution in general relativity.
\newblock {\em Phys. Rev.}, 153:1359--1363, Jan 1967.

\end{thebibliography}
\bibliographystyle{unsrt}

\end{document}